\documentclass[a4paper,11pt]{article}
\pdfoutput=1 

\usepackage{style/jinstpub} 
\usepackage[utf8]{inputenc}

\newcommand{\centim} {\ensuremath{ \,{\mathrm{cm}}}}
\newcommand{\milim}  {\ensuremath{ \,{\mathrm{mm}}}}

\newcommand{\second} {\ensuremath{ \,{\mathrm{s}}}}
\newcommand{\hour}   {\ensuremath{ \,{\mathrm{h}}}}

\newcommand{\KeV} {\ensuremath{ \,{\mathrm{KeV}}}}

\newcommand{\fbi}    {\ensuremath{ \,{\mathrm{fb^{-1}}}}}
\newcommand{\uinstlumi}    {\ensuremath{ \,\centim^{-2}\second^{-1}}}

\newcommand{\mGy}    {\ensuremath{ \,{\mathrm{mGy}}}}

\title{\boldmath Study of the effects of radiation on the CMS Drift Tubes Muon Detector for the HL-LHC}

\author[a]{G.~Abbiendi}
\author[b]{J.~Alcaraz Maestre}
\author[b]{A.~\'Alvarez Fern\'andez}
\author[c]{B.~\'Alvarez Gonz\'alez}
\author[d,e]{N.~Amapane}
\author[b]{I.~Bachiller}
\author[b]{J.~M.~Barcala}
\author[f]{L.~Barcellan}
\author[a,g]{C.~Battilana}
\author[f]{M.~Bellato}
% ...

\affiliation[a]{INFN Sezione di Bologna, Bologna, Italy}
\affiliation[b]{Centro de Investigaciones Energ\'eticas Medioambientales y Tecnol\'ogicas (CIEMAT), Madrid, Spain}
\affiliation[c]{Instituto Universitario de Ciencias y Tecnolog\'ias Espaciales de Asturias (ICTEA), Universidad de Oviedo, Oviedo, Spain}
\affiliation[d]{INFN Sezione di Torino, Torino, Italy}
\affiliation[e]{Università di Torino, Torino, Italy}
\affiliation[f]{INFN Sezione di Padova; Università di Padova, Padova, Italy}
\affiliation[g]{Università di Bologna, Bologna, Italy}
\affiliation[h]{Wigner Research Centre for Physics, Budapest, Hungary}
\affiliation[i]{Institute of Nuclear Research ATOMKI, Debrecen, Hungary}
\affiliation[j]{Instituto de Física de Cantabria (IFCA), CSIC-Universidad de Cantabria, Santander, Spain}
\affiliation[k]{Universidad Autónoma de Madrid (UAM), Madrid, Spain}
\affiliation[l]{RWTH Aachen University, III. Physikalisches Institut A, Aachen, Germany}
\affiliation[m]{Laboratori Nazionali di Legnaro dell'INFN, Legnaro, Italy}
\affiliation[n]{Institute of Physics, University of Debrecen, Debrecen, Hungary}
\affiliation[o]{Italian National Agency for New Technologies, Energy and sustainable economic development, Bologna, Italy}
\affiliation[p]{Now at CERN}
\affiliation[q]{Now at RWTH Aachen University}
\affiliation[\dag]{Deceased}

\emailAdd{gonzalezisidro@uniovi.es}

\abstract{The CMS drift tubes (DT) muon detector, built for withstanding the LHC expected integrated and instantaneous luminosities, will be used also in the High Luminosity LHC (HL-LHC) at a 5 times larger instantaneous luminosity and, consequently, much higher levels of radiation, reaching about 10 times the LHC integrated luminosity. Initial irradiation tests of a spare DT chamber at the CERN gamma irradiation facility (GIF++), at large ($\sim$O(100)) acceleration factor, showed ageing effects resulting in a degradation of the DT cell performance. However, full CMS simulations have shown almost no impact in the muon reconstruction efficiency over the full barrel acceptance and for the full integrated luminosity. A second spare DT chamber was moved inside the GIF++ bunker in October 2017. The chamber was being irradiated at lower acceleration factors, and only 2 out of the 12 layers of the chamber were switched at working voltage when the radioactive source was active, being the other layers in standby. In this way the other non-aged layers are used as reference and as a precise and unbiased telescope of muon tracks for the efficiency computation of the aged layers of the chamber, when set at working voltage for measurements. An integrated dose equivalent to two times the expected integrated luminosity of the HL-LHC run has been absorbed by this second spare DT chamber and the final impact on the muon reconstruction efficiency is under study. Direct inspection of some extracted aged anode wires presented a melted resistive deposition of materials. Investigation on the outgassing of cell materials and of the gas components used at the GIF++ are underway. Strategies to mitigate the ageing effects are also being developed. From the long irradiation measurements of the second spare DT chamber, the effects of radiation in the performance of the DTs expected during the HL-LHC run will be presented.}

\collaboration[c]{on behalf of the CMS Muon group}

\proceeding{21\textsuperscript{st} International Workshop on Radiation Imaging Detectors - iWoRiD 2019 \\
 7--12 July 2019\\
 Orthodox Academy of Crete (OAC), Kolymbari, Crete, Greece}

\begin{document}
\maketitle
\flushbottom

\section{Introduction}
\label{sec:intro}
In 2026 the LHC will change its running conditions and enter its phase II: the so-called High Luminosity LHC (HL-LHC)~\cite{HLLHC}. In this phase the foreseen instantaneous luminosity ($\sim 5 \cdot 10^{34} \uinstlumi$) will be a factor 5/7 higher than the current one at LHC and the accumulated luminosity by the end of the data taking around 2035 ($3000 \fbi$) will be 10 times larger than the total delivered luminosity in phase I. 

All the detectors operating at the HL-LHC need to take the appropriate actions to ensure they can work under these stringent radiation conditions. The Compact Muon Solenoid (CMS) detector~\cite{CMS} is currently operating at the LHC at a great level of performance. However by the time of the HL-LHC the different subdetectors will need to work in an environment they were not originally designed for. Many of them will overcome important improvements and some components will be replaced. The Drift Tubes (DT) subdetector is an essential part of the CMS muon spectrometer~\cite{DT} and plays a key role in the identification of muons. An assessment of its capabilities to perform correctly under high levels of radiation and during a long period is mandatory.

In order to estimate the effects due to the higher instantaneous and integrated dose expected during the HL-LHC in the DT chambers some tests are being conducted in the CERN Gamma Irradiation Facility (GIF++)~\cite{GIF}.

\subsection{The CMS Drift Tubes subdetector (DT)}
\begin{figure}[htbp]
\centering 
\includegraphics[width=.8\textwidth]{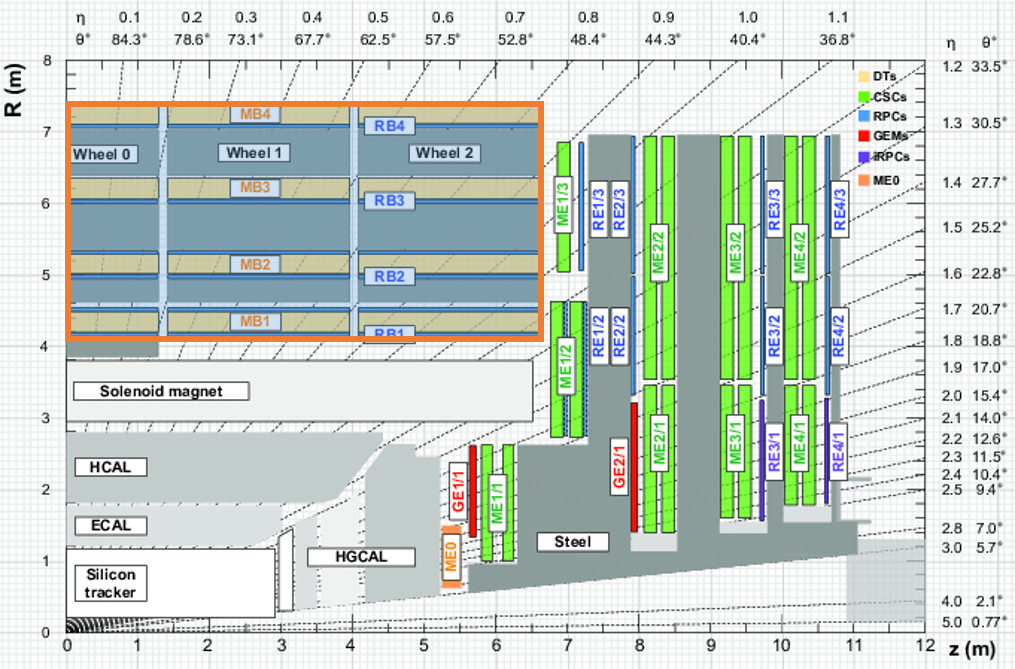}
\caption{\label{fig:CMS}  Schematic longitudinal view of one quarter of the CMS detector. The orange square highlights the region where the DT chambers (MB1, MB2, MB3 and MB4) are located.}
\end{figure}

The DT subdetector consists of 250 chambers located in 5 wheels covering the barrel region of the CMS detector inside $|\eta < 1.2|$. The DTs are responsible for muon identification and momentum measurement over a wide range of values. The DT system also provides a robust standalone muon trigger with precise bunch crossing assignment. A schematic view of CMS is pictured in figure~\ref{fig:CMS}.

\begin{figure}[htbp]
\centering 
\includegraphics[width=.33\textwidth]{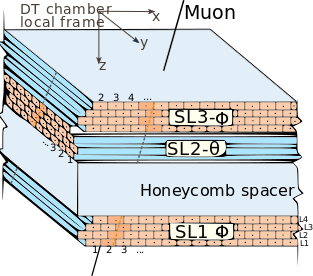}
\qquad
\includegraphics[width=.57\textwidth]{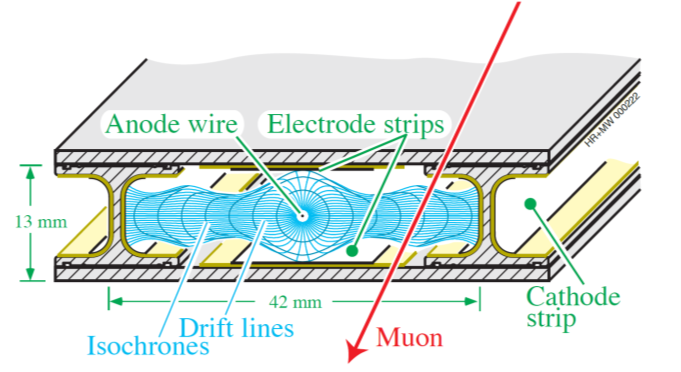}
\caption{\label{fig:cell} Left: Structure of a DT Chamber with three superlayers composed of 4 layers each of drift cells. Right: Layout of a drift cell showing the electric field lines in the gas volume.}
\end{figure}

Each DT chamber is a glued structure of about 700 drift cells. The basic detector element is a rectangular drift cell delimited by aluminium beams on the sides, where cathodes are located, and aluminium plates on the top and bottom. Aluminum strip electrodes shape the electric field to achieve constant drift velocity along the cell as shown in Figure~\ref{fig:cell}. The drift cells are grouped in three fully independent units (superlayers, SL) with respect to gas circulation, High Voltage (HV) distribution and Front-End amplifiers (FEB). Each superlayer comprises four staggered layers of parallel 50 to 100 drift cells.

\subsection{The Gamma Irradiation Facility (GIF++)}
GIF++ is located in the North Area of the CERN Super Proton Synchrotron (SPS). The facility is a unique place where high energy charged particle beams (mainly muons) from the SPS are combined with a flux of photons from a 13.9~TBq $^{137}$Cs source. The set-ups are organized in two areas inside the bunker: upstream and downstream corresponding to the side of the source closer to or further from the beam starting point. A filter system permits attenuating the photon rate independently in the two areas in several steps to reach attenuation factors of several orders of magnitude ($\sim10^4 - 10^5$). This high activity source allows to accumulate doses equivalent to HL-LHC experimental conditions in a reasonable amount of time. 

\subsection{Effects of radiation on gaseous detectors}

The performance degradation of irradiated gaseous detectors is well documented in the literature~\cite{gasirradiated}. A first study of the behaviour of a DT chamber under irradiation was carried out in 2015 and 2016. A spare MB1 DT chamber was placed in the downstream area of GIF++ and it was irradiated during a few months at a dose rate about 100 times higher than the expected one during HL-LHC. The chamber showed a fast gas gain drop during the first few weeks of irradiation~\cite{PhaseIITDR}. The electron avalanche conditions, together with the ionizing radiation, enable chemical reactions that are able to break covalent bonds and generate free radicals that can recombine forming deposits on the surface of the anode wire. The degradation of the detector is typically caused by a polymerization process that produces a hydrocarbon and silicon deposit on the anode. Some irradiated wires were inspected using microscopy techniques and a carbon and silicon coating having large resistivity was observed.

\section{Experimental setup}
A more systematic experiment was conducted using a spare DT MB2 chamber that was introduced in the GIF++ bunker in September 2017. The chamber was irradiated at a low accelerating factor ($\sim 10$ times the expected dose rate at HL-LHC) in two campaigns that finished in January 2019. The chamber was situated at about 4 m from the source, in the downstream part of the GIF++ bunker. The chamber was standing in vertical position with the SL1 facing the source. 

The chamber was continuously monitored and the detection performance was checked weekly under different operational and background conditions. Data with a muon beam was also recorded in October 2017 and July 2018. 

Since the ageing effects are expected to happen only when the potential difference in the drift cells is high, it was decided to have only two layers (L1 and L4) in SL1 with anode HV at working voltage (3550~V) when the radioactive source was active. The rest of the chamber was kept idle at 1900~V so the gas gain is negligible. This arrangement preserves the other layers so they can be used as a precise and unbiased telescope for muon tracks.

\subsection{Dose estimation}
While at GIF++ the amount of charge absorbed by the detector is proportional to the radiation dose caused by  662 \KeV~(50\%) and low energy photons, the background radiation in CMS is mainly composed by low energy neutrons. A permanent REMUS dosimeter in the GIF++ bunker monitors the dose rate. Detailed measurements with a portable dosimeter were used to extrapolate from the REMUS position to the real location of the chamber. The total accumulated dose can be computed at any time by integrating the dose rate over the periods where the aged layers were kept on.

\begin{figure}[htbp]
\centering 
\includegraphics[width=.8\textwidth]{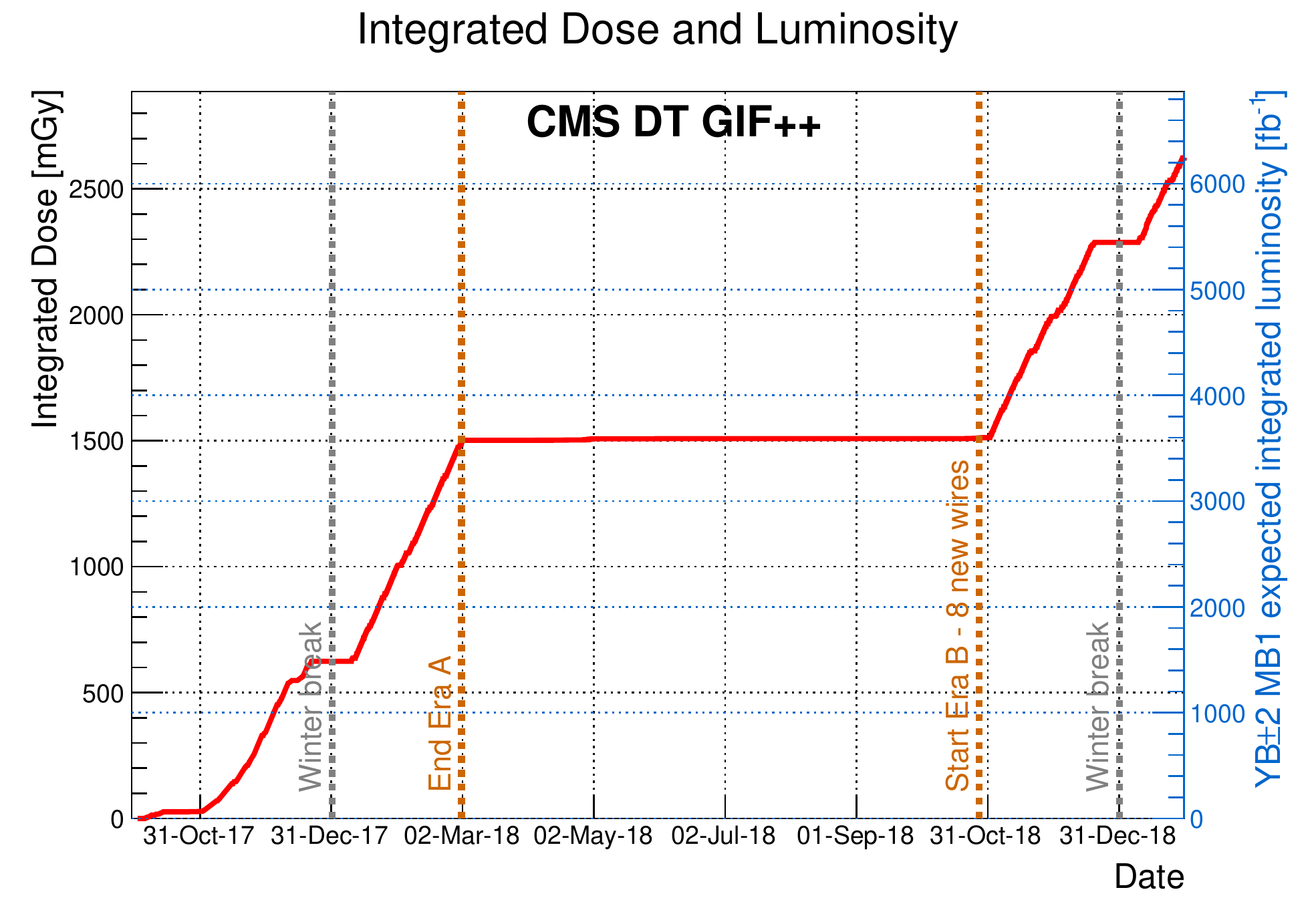}
\caption{\label{fig:datataking} Integrated dose over the irradiated layers (L1 and L4) in SL1. Only the periods when they were on are considered in the computation. The data taking is divided into two periods:  the first, called era A, started in October 2017 and ended in April 2018, after submitting the chamber to an amount of radiation equivalent to that expected for the full HL-LHC run. After that period, the DT chamber was open and a few wires were extracted for inspection. The era B started at the end of October 2018 and continued until February 2019.}
\end{figure}

At CMS, the integrated dose received by any of the DT stations is proportional to the integrated luminosity. A conversion factor from integrated dose at GIF++ to the expected integrated luminosity at CMS was derived for the most exposed chamber (MB1 chambers in the external wheels of the CMS detector): $1 \fbi = 0.42 \mGy$. Furthermore, the  effect of the background rate also affects the performance of the detector. The muon hit efficiency is degraded by the presence of a background of particles that can affect the signal collected by the anode wires or introduce background noise to the measurements. The effect of the background rate at GIF++ is proportional to the photon dose rate from the source. For the CMS DT system, the amount of background rate for each DT station depends on the instantaneous luminosity given by the LHC. The relation between these two quantities is $10^{34} \uinstlumi = 0.0109 \mGy\hour^{-1}$. Figure~\ref{fig:datataking} shows the evolution of the accumulated dose and its equivalent expected luminosity under HL-LHC conditions over time.

\subsection{Hit efficiency}

The DT hit detection efficiency plays a key role in the ability of the CMS detector to identify, trigger and reconstruct muons. A big effort was put in characterizing this quantity as the cells aged. The efficiency to detect a single hit in a cell of a layer was defined and measured as the ratio between the number of detected and expected hits. The position of expected hits was determined using as probes sets of well reconstructed track segments with associated hits in at least 4 layers in SL3 and at least 1 layer in SL1. The intersection of this track segment with the layer under study determined the position, therefore the cell, where a hit was expected: the cell was considered efficient if a hit was found within it. 


\section{Results}

\begin{figure}[htbp]
\centering 
\includegraphics[width=.8\textwidth]{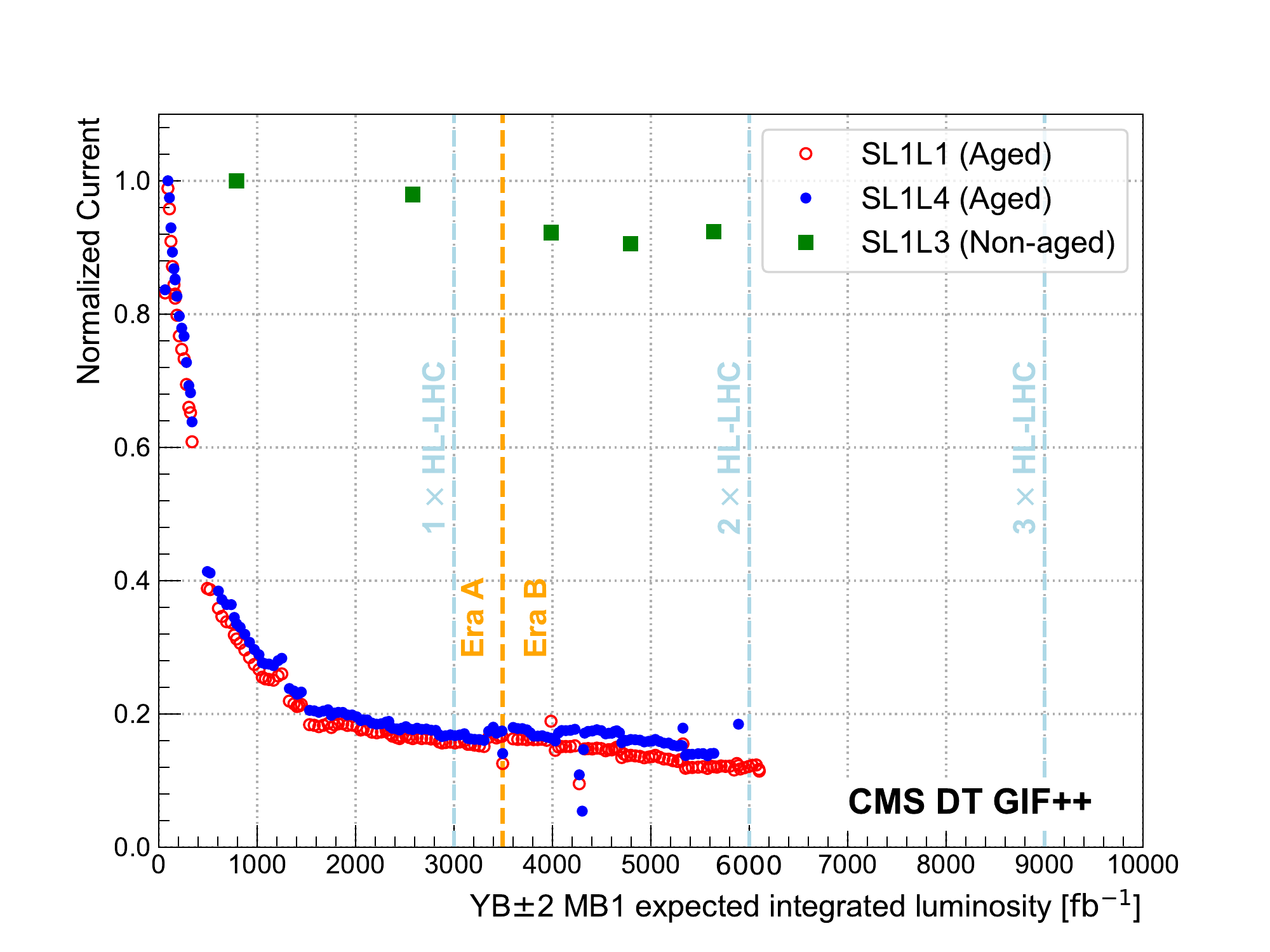}
\caption{\label{fig:normcurrent} Normalized Current for aged (SL1L1 and SL1L4) and non-aged (SL1L3) layers as a function of the integrated luminosity. Values are scaled with respect to the initial value in SL1L3. Each point corresponds to the mean current in a day.}
\end{figure}

A first hint of the ageing can be inferred from the evolution of the current recorded in the anode in the presence of constant background rate. We define the normalized current as the ratio between the instantaneous current and the instantaneous dose which is very closely related to the gain in the chamber. Since the DT chambers are not completely hermetic they are affected by the pressure in the bunker and the gain changes accordingly. The pressure in GIF++ is recorded continuously and used to correct the current measured in the anode. Figure~\ref{fig:normcurrent} shows the evolution of the normalized current for aged (SL1L1 and SL1L4) and non-aged (SL1L3) layers at 3550 V as a function of the integrated luminosity. A big drop at the beginning and a tendency to a plateau after about 1500~\fbi is observed. This suggests that the polymerization process responsible for the loss of gain happens faster when the wire has less coating and it saturates at higher values of integrated dose.

\begin{figure}[htbp]
\centering 
\includegraphics[width=.49\textwidth]{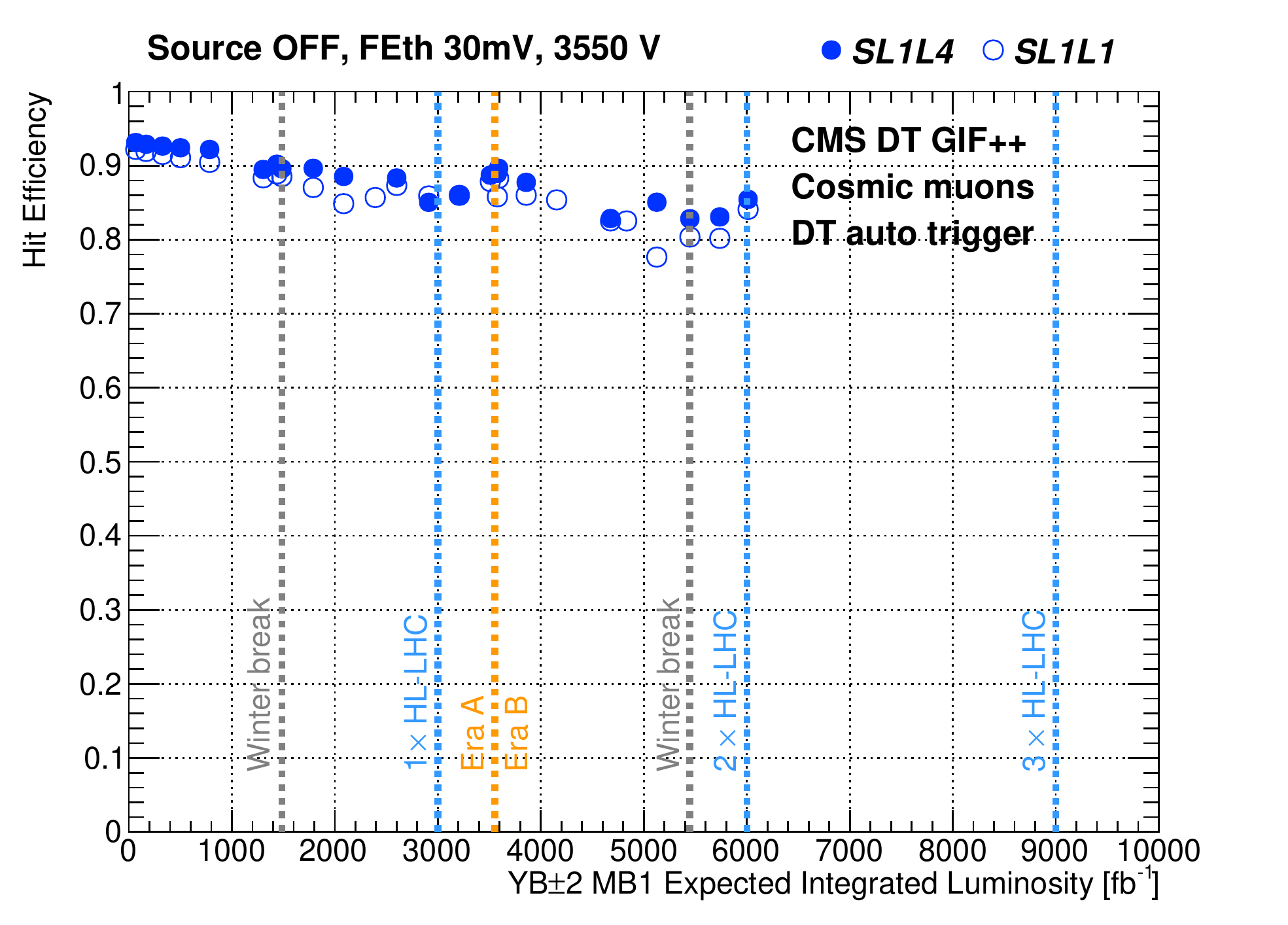}
\hfill
\includegraphics[width=.49\textwidth]{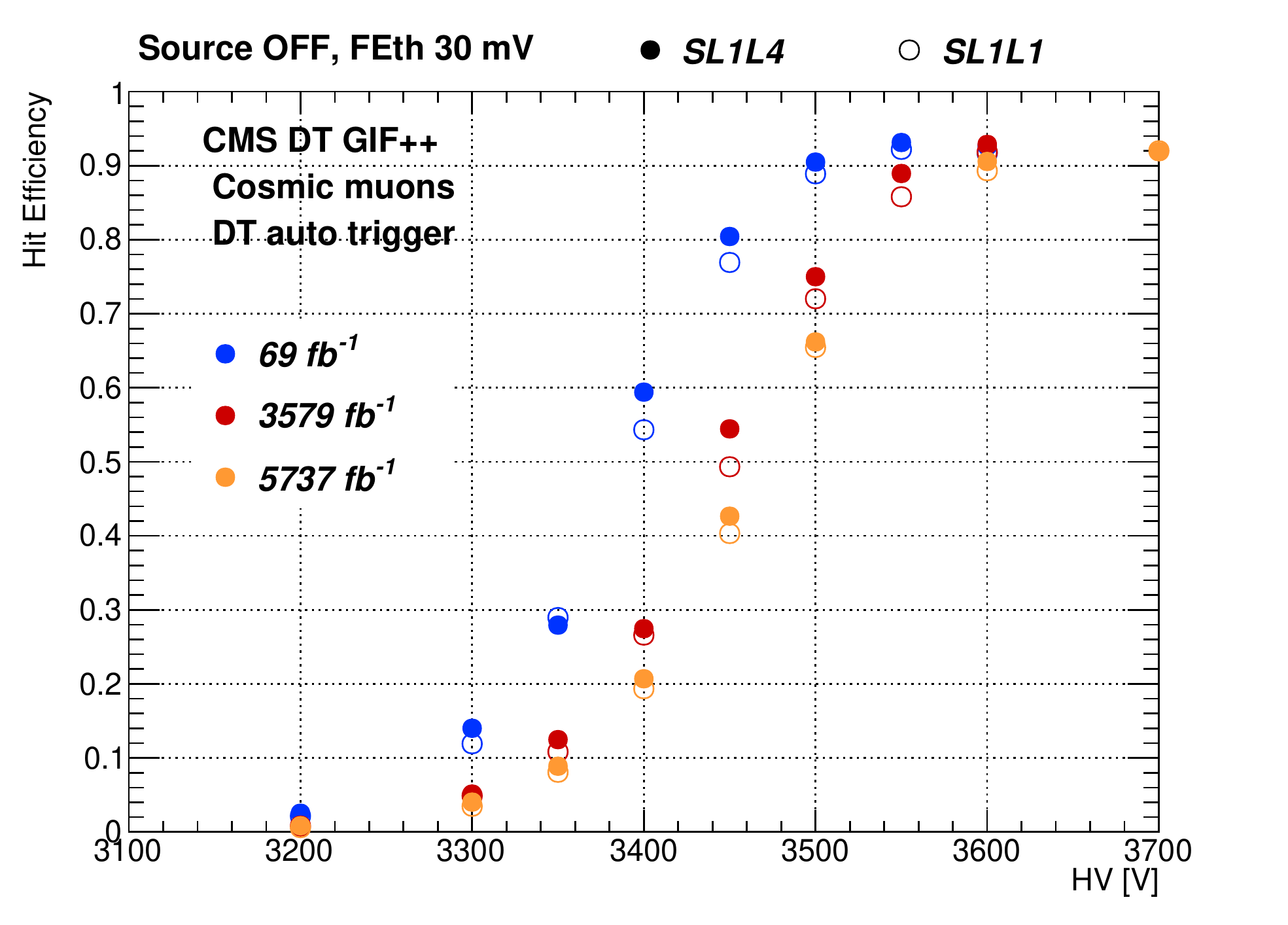}
\caption{\label{fig:effCosmic} Hit efficiency for the aged layers (SL1L1 and SL1L4) using cosmic muons as a function of integrated luminosity (left) and as a function of the wire HV for different integrated luminosities (right). The integrated luminosity corresponds to that expected for the MB1 chambers in the external wheels of the CMS detector (YB$\pm2$) during the HL-LHC.}
\end{figure}

For most of the data taking cosmic muons were used to measure the efficiency at different operational conditions. Cosmic muons were recorded using the DT auto trigger requiring a track in both projections on just SL2 and SL3 of the DT chamber. Figure~\ref{fig:effCosmic} (left) shows the evolution of the hit efficiency in the aged layers using cosmic muons as a function of the equivalent integrated luminosity for the most exposed chambers. A moderate efficiency loss is observed in both aged layers of about 10\% for an accumulated irradiation equivalent to twice the expected integrated luminosity at the HL-LHC. In figure~\ref{fig:effCosmic} (right) this effect can be seen for different values of the anode HV in three moments of the irradiation period. 

\begin{figure}[htbp]
\centering 
\includegraphics[width=.49\textwidth]{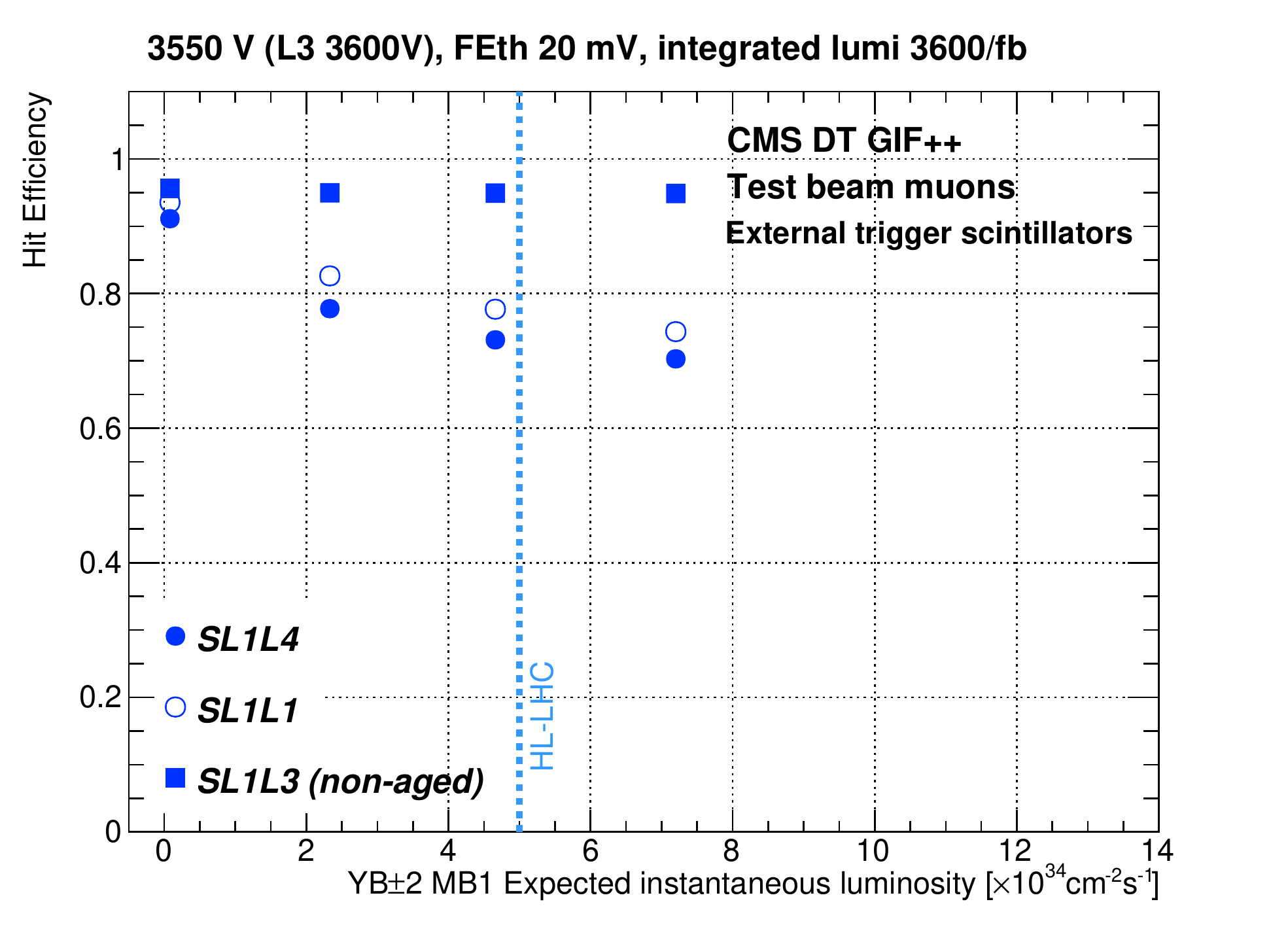}
\hfill
\includegraphics[width=.49\textwidth]{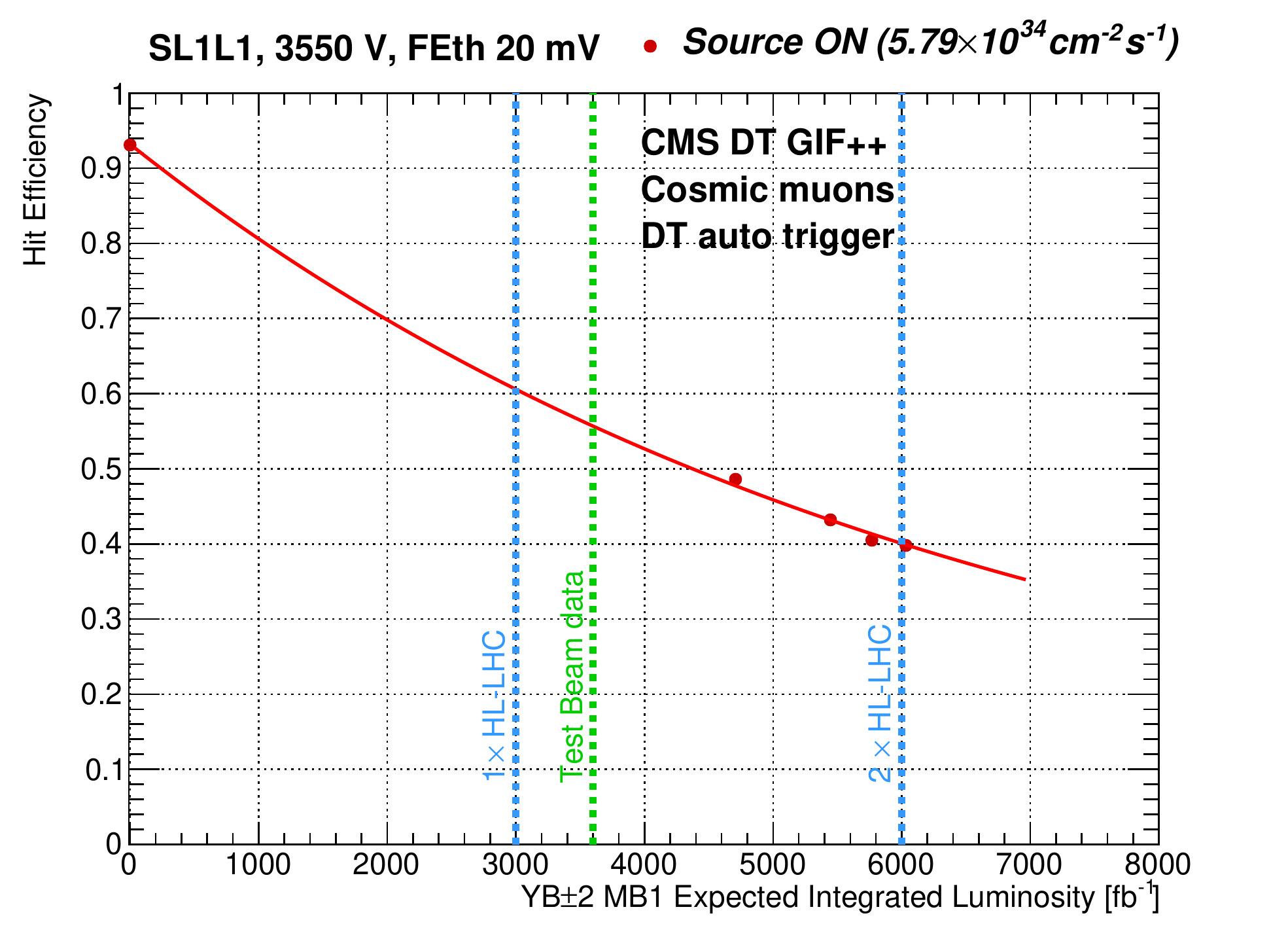}
\caption{\label{fig:effMuon} Left: Hit efficiency for test beam muons as a function of instantaneous dose for aged layers (SL1L1 and SL1L4) and a non-aged layer (SL1L3). Data taken after an accumulated dose equivalent to an HL-LHC integrated luminosity of 3600~\fbi. Right: Hit efficiency for cosmic muons as a function of integrated luminosity for the aged layer SL1L1 at a background rate corresponding to a instantaneous luminosity of $5.8 \cdot 10^{34} \uinstlumi$). The data is fitted using an exponential model ($\varepsilon_{hit} = 0.053 + 0.879 \cdot e^{-1.546 \cdot 10^{-4}\mathcal{L}_{int}}$).}
\end{figure}

To have a more realistic scenario data was taken with high momentum muons together with the source in place using different filters that would mimic different background conditions. GIF++ external scintillators were used to trigger these muons.  Figure~\ref{fig:effMuon} (left) shows the dependence of the hit efficiency with the instantaneous luminosity for the aged layers when the chamber had accumulated a dose a bit larger than that expected at the HL-LHC. The hit efficiency degraded about 20\% in the presence of background radiation. The non irradiated layers show no drop. 

To cope with the various uncertainties in the final HL-LHC conditions and the data taking, it was decided to consider a safety factor of 2 on both the expected instantaneous and integrated luminosity in the most exposed chambers. To extrapolate the hit efficiency in the presence of a background rate equivalent to $\cdot 10^{35} \uinstlumi$ an exponential fit of the data in figure~\ref{fig:effMuon} (left) is performed and used averaging over the two curves.

Unfortunately no beam data could be taken at an integrated dose equivalent to 6000~\fbi. A few short measures with cosmic muons were performed with the source on to minimize ageing on the reference layers. Figure~\ref{fig:effMuon} (right) shows the evolution of the efficiency using cosmic muons for a background rate slightly higher than the one expected at the HL-LHC. The fit was later used to estimate the efficiency at an expected integrated luminosity of 6000~\fbi.

\begin{figure}[htbp]
\centering 
\includegraphics[width=.73\textwidth]{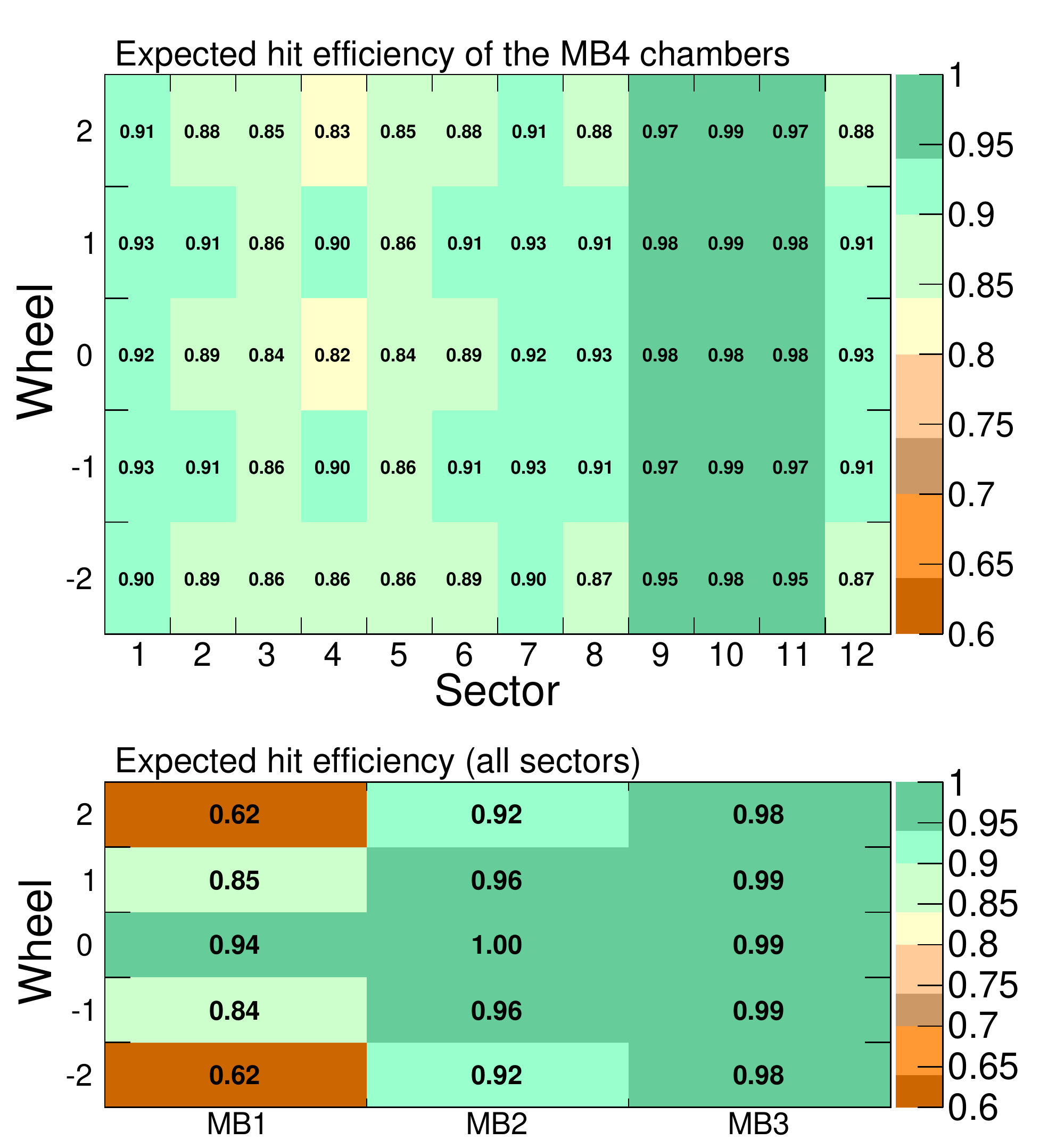}
\caption{\label{fig:effsfinal} Estimated hit efficiencies at the end of the HL-LHC for all the DT chambers of the CMS muon system, MB4 chambers (upper) and for MB1, MB2 and MB3 (lower). These efficiencies have been estimated considering a safety factor of 2 for the expected HL-LHC background rate ($10^{35} \uinstlumi$) and a safety factor of 2 for the expected integrated luminosity (6000~\fbi) to obtain the expected hit efficiency for the MB1 chambers in wheels ±2 and extrapolating to the rest of the CMS muon system using the expected integrated charge at the end of HL-LHC.}
\end{figure}

All these measurements summarize the characterization of the expected ageing of a DT chamber at the HL-LHC. In particular, we chose to study the effect of the ageing in terms of expected instantaneous and integrated luminosity at the HL-LHC for a MB1 chamber at wheels ±2, which will be the most irradiated DT chambers. To extrapolate the measured efficiency to the rest of the DT chambers of
the CMS muon system, the expected background rate and integrated luminosity are
obtained from the measurement of the integrated charge during the 2018 data taking at CMS for the 250 DT stations, correcting by the expected changes in the detector before the starting of the HL-LHC run. The values obtained are detailed in Figure~\ref{fig:effsfinal}.

\section{Conclusions}
The HL-LHC will heavily increase the data available in particle physics at the cost of putting the detectors in a harsher environment where the radiation hitting the detectors will be largely increased. Drift cells are known to loose performance as they accumulate radiation. On the other hand, the additional background rate coming with the increase of instantaneous luminosity has the capability of further affecting the performance of aged detectors. A big effort has been done by the CMS Muon Project to asses the capability of the DT system to work under these conditions.

The data collected by irradiating spare DT chambers at GIF++ has provided very valuable data to characterise their response throughout the HL-LHC operations. A pessimistic extrapolation based on the data collected at the CERN GIF++ facility shows a drop in the hit efficiency below 25\% at the end of the HL-LHC for the most exposed chambers while the big majority of the system stays well above 90\%. Very preliminary muon trigger and reconstruction studies show a mild localized effect in the overlap region due to the redundancy of the system.

Mitigation strategies have been implemented already to reduce the possibility of ageing. Since the majority of the radiation that affects the chambers comes from the scattering on the walls and the activated materials in the CMS experimental cavern, a 7~\milim~Lead and 30-90~\centim~5\% borated polyethylene shielding is being installed on top of the most exposed DT chambers. At the same time the operational HV of the chambers has been reduced keeping the same performance of the detector.

A dynamic management of the  operational conditions (HV, FE Threshold) can be used during the HL-LHC to improve the lifespan of the muon detectors. For example, a lower HV could be used at the beginning to reduce ageing. As signs of ageing appear the HV can be increased to get back again into the plateau (see Figure~\ref{fig:effCosmic} right).

Finally, investigation of the effect of additional components to current gas mixtures that may eventually moderate the coating deposition is planned.


\end{document}